\documentclass[final,5p,times]{elsarticle}
\usepackage{amsmath,bm,amssymb,textcase,mathrsfs,bbm,amscd,xspace}           % Mathematics packages
\usepackage{graphicx}                  % Figures and graphics
\usepackage{listings}                  % Code listings
\usepackage{cuted}                  % Color for code
\usepackage{hyperref}                  % Hyperlinks
\usepackage{minted}
\usepackage{graphicx}% Include figure files
\usepackage{bm}% bold math
\usepackage{lipsum}
\usepackage{color,soul}
\usepackage{dsfont}
\usepackage{makecell}
\usepackage{youngtab}
\usepackage{dsfont}
\usepackage{cuted,booktabs,longtable}
\usepackage{float}

\newcommand{\suf}{\mbox{${\rm{SU_{ST}}(4)}$}}

\newcommand {\arr}{\begin{array}}
\newcommand {\rra}{\end{array}}

\newcommand {\eq}{\begin{equation}}
\newcommand {\qe}{\end{equation}}
\newcommand {\eqa}{\begin{eqnarray}}
\newcommand {\qea}{\end{eqnarray}}

\newcommand{\usum}{\sum\limits}
\newcommand*{\Resize}[2]{\resizebox{#1}{!}{$#2$}}%
\newcommand{\msc}[1]{\mathscr{#1}}

\begin{document}

%-------------------------
% Front Matter
%-------------------------
\begin{frontmatter}
\title{Automated Calculation of Spin–Isospin Branching Rules for SU(4) Irreps}
\author[inst1]{S. Quintero \corref{cor1}}
\author[inst1]{R.\ Henao\corref{cor1}}
\author[inst1]{J.\ P.\ Valencia\corref{cor1}}
\cortext[cor1]{Corresponding author: ' patricio.valencia@udea.edu.co}
\address[inst1]{Institute of Physics, Universidad de Antioquia, Medell\'in, Colombia}

\begin{abstract}
The open-source Python package, \texttt{su4-branching}, is introduced for the derivation of comprehensive spin $S$ and isospin $T$ branching rules for any SU(4) irreducible representation.The Wigner supermultiplet scheme in nuclear and hadronic physics is based on SU(4) symmetry. However, the community does not have easy access to practical calculations of branching rules for any irreps.Our implementation combines group-theoretical methods with a notebook interface that is easy to use. This lets researchers look at large and complicated SU(4) irreps and check their work.The software produces tables, CSV files, and visual summaries, and it has been tested against both classic and modern reference results. This work enables group-structure investigations in nuclear modeling, particle physics, and quantum chemistry.
\end{abstract}


\end{frontmatter}

%-------------------------
% Program Summary
%-------------------------
%-------------------------
% Program Summary
%-------------------------
\section*{Program Summary}
\begin{small}
\begin{tabular}{@{}p{3cm}p{11cm}@{}}
\textbf{Program Title:}      & su4branching \\
\textbf{Licensing:}          & MIT License \\
\textbf{Programming language:}& Python 3.8+ \\
\textbf{Dependencies:}       & NumPy, SymPy, pandas, matplotlib \\
  \textbf{Operating system:}   & Cross‐platform\\
  &(Linux, macOS, Windows) \\
  \textbf{Keywords:}           & SU(4), branching rules, spin, isospin,\\
                               &group theory \\
  \textbf{Nature of problem:}  & Automated decomposition of\\
                             &high‐dimensional SU(4) irreps\\
  &into spin–isospin multiplets. \\
  \textbf{Solution method:}    & The solution use the Racah \\
  &formula ~\cite{racah1949,lopez1990}.\, \\
  &Outputs in LaTeX, CSV, and plots. \\
\textbf{Restrictions:}        & test up-to 24 particles \\
  \textbf{Unusual features:}   & cumulative dimension checks,\\
                             &Jupyter notebook integration,\\
  &aplication $U(6)\otimes SU(4)$, $U(10)\otimes SU(4)$.
\end{tabular}
\end{small}

\section{Introduction}
Symmetry considerations play a central role in theoretical and computational physics. In nuclear
structure, the SU(4) group encapsulates combined spin and isospin degrees of freedom, giving
rise to the Wigner supermultiplet scheme for light nuclei. Analytical determination of all possible
spin-isospin (S, T) multiplets inside an SU(4) irreducible representation are essential for shell model calculations, beta decay studies, quantum optics, quantum chemistry, and beyond. However, for higher-dimensional irreps or general Young tableau shapes, producing these branching rules by hand becomes impractical and error-prone. Despite the theoretical background developed since the foundational work of Wigner and
Hecht–Pang and recent advances in the calculation of \suf\, Wigner coefficients ~\cite{wigner1937,hecht1969,pan2024}, no general-purpose, ready-to-use computational tool has been freely available for this task. To address this need, we introduce a Python-based code, \texttt{su4-branching}, which
automates the calculation of the complete set of branching rules (S, T) and their multiplicities for
any input of \suf\, irrep. Our implementation enables systematic exploration, benchmarking, and
validation for research in nuclear theory, group-based quantum many-body modeling, and
related fields.
\section{Importance of Branching Rules}
The branching rules of \suf\, $\downarrow$ SU$_S$(2) $\otimes$ SU$_T$(2) play a fundamental role in the application of the Wigner supermultiplet scheme to nuclear structure theory ~\cite{hecht1969,isacker1995,isacker1999,isacker2016,draayer2024,kotasahu2024}. These rules are essential for interconnected aspects of nuclear structure calculations.
\subsection{ Calculation of SU(4) Wigner and Racah coefficients}
The branching rules are the basis for figuring out the reduced Wigner coefficients and Racah coefficients for the \suf\,  group chain~\cite{wigner1937,hecht1969,pan2024,dang2024}.\, You can break these coefficients down into smaller ones.\, These coefficients, along with the regular SU(2) Wigner coefficients, enable a systematic understanding of how spin-isospin coupling operates in nuclear matrix elements. You can write the general form of a reduced matrix element as
\begin{equation}
\begin{array}{l}
\langle(\gamma S T) || \mathcal{O}^{(\kappa)} || (\gamma' S' T') \rangle = \usum_{\alpha} C_{\alpha}(\gamma, \gamma', \kappa)W^{(\alpha)}(S, T, S', T').
  \end{array}
\end{equation}
In this equation, $\mathcal{O}^{(\kappa)}$ is a tensor operator of rank $\kappa$ in SU(4), $C_{\alpha}$ are the reduced SU(4) coupling coefficients that obey the branching rules, and $W^{(\alpha)}$ are the standard SU(2) $\otimes$ SU(2) Wigner coefficients. To find the fractional parentage coefficients (cfp) and matrix elements of one- and two-body operators in the supermultiplet scheme~\cite{hecht1975,quesne1976,partensky1978,braun1978}, you need to calculate these coefficients.
\subsection{Orthonormality requirements and basis construction}
The branching rules guarantee the formation of orthonormal basis vectors for irreducible representations~\cite{hecht1969}. It is important to do proper orthonormalization to get physically meaningful matrix elements and to keep the many-body wave function expansion consistent. The orthonormality conditions for \suf\, basis states must be met
\begin{equation}
\langle (\gamma \beta \nu) S T \alpha | (\gamma' \beta' \nu') S' T' \alpha' \rangle = \delta_{\gamma\gamma'} \delta_{\beta\beta'} \delta_{\nu\nu'} \delta_{SS'} \delta_{TT'} \delta_{\alpha\alpha'},
\end{equation}
In this context, $\alpha$ represents any additional quantum numbers required to address multiplicities. These orthonormality conditions guarantee that the reduced matrix elements follow the Wigner-Eckart theorem in the \suf\, framework and that nuclear wave functions expressed in the supermultiplet scheme are properly normalized.
\subsection{Classification of nuclear states}
The branching rules allow for the systematic categorization of nuclear states based on their spin-isospin symmetry characteristics. For instance, in ds-shell nuclei, nuclear states adhere to the group-theoretical reduction chain.
\begin{strip}
\begin{equation}
\label{eq:dsqueme}
\begin{array}{ccccccccc}
\mathrm{U}(24)\supset&\Biggl(\mathrm{U}_{L}(6)  \supset&
\mathrm{SU_{L}}(3)
\supset&\mathrm{SO}(3) & \Biggl)&\otimes&\Biggl(\mathrm{SU}_{S,T}(4) \supset &\Bigg(\mathrm{SU}(2)_{S} \otimes \mathrm{SU}(2)_{T} \Bigg)\Biggl)&\\
\downarrow&\downarrow&\downarrow&\downarrow&&&\downarrow&\downarrow&\\
\Biggl|\left[1^{N_{\mathrm{ds}}}\right]&\{f\}&
\chi(\lambda \,\mu )&k L&& & (\gamma\,\beta\,\nu)& \phi\left(\,\,S\,T\right)&
\,\,\Biggl\rangle
\end{array},
\end{equation}
\end{strip}
where $N_{\mathrm{ds}}$ stands for the number of nucleons in the ds-shell and $\{f\}$ is the Young diagram that shows the irreducible representation of $\mathrm{U}_{L}(6)$, In Dynkin notation, $(\lambda, \mu)$ marks the $\mathrm{SU}_{L}(3)$ irreducible representation, $L$ marks the angular momentum quantum number, and $(\gamma, \beta, \nu)$ marks the \suf\, irreducible representation. The quantum numbers $(S,T)$ tell us the total spin and isospin. The label $\chi(k)$ differentiates repeated instances of the irreducible representation $(\lambda, \mu)(L)$ within a specific $\{f\} ((\lambda, \mu))$, whereas $\phi$ clarifies the multiplicity of $(S,T)$ states within the \suf\, irreducible representation $(\gamma, \beta, \nu)$.
\subsection{Understanding of approximate symmetries}
Branching rules indicate that \suf\, symmetry, although affected in realistic nuclei by spin-orbit coupling and other factors~\cite{isacker1995}, still captures a significant portion (up to 90\%) of the spin-isospin structure of nuclear wave functions. This newly discovered approximate symmetry enhances our understanding of nuclear binding energies, weak transition strengths, and the reduced likelihood of transitions between states belonging to different \suf\, irreducible representations. The extent of deviation from exact \suf\, symmetry can be assessed by examining symmetry-breaking matrix elements.
\begin{equation}
\Delta E_{\suf\,} = \langle \Psi_{\suf\,} | H - H_{\suf\,} | \Psi_{\suf\,} \rangle,
\end{equation}
In this expression, \( H \) denotes the complete nuclear Hamiltonian, whereas \( H_{\suf} \) represents its \suf-invariant component. Comprehending these deviations via the appropriate use of branching rules is crucial for linking the idealized symmetry framework to practical nuclear structure computations.
\section{Branching Method}
We use Racah's tensor-contraction method~\cite{racah1949,lopez1990} to discover the branching rules and multiplicities for the reduction of \suf\, irreducible representations, which are labeled by their Young-diagram row lengths $(f_1, f_2, f_3,0)$ to $\mathrm{SU}(2)_S \otimes \mathrm{SU}(2)_T$ multiplets. This method provides a quick combinatorial algorithm that does not require explicitly diagonalizing a matrix. It provides the multiplicity $\msc{N}_{ST}(f_1, f_2, f_3,0)$ for each $(S,T)$ pair using closed-form formulas utilizing greatest-integer functions. Racah's technique has a significant computational advantage for large irreducible representations. This advantage is because scaling in the Young-diagram dimensions is linear rather than cubic, as with projection-matrix approaches.  The branching of a \suf\, irrep into $\rm{SU(2)_S} \otimes \rm{SU(2)_T}$ multiplets is given by
\begin{equation}
\label{eq:racah-a}
(f_1\, f_2\, f_3)
\;\downarrow\;
\bigoplus_{S,T}
\msc{N}_{S\,T}(f_1\, f_2\, f_3)\,(S\,T),
\end{equation}
 where $\msc{N}_{ST}(f_1 f_2 f_3)$ is the number of times the $(S\,T)$ representation occurs.\, The Racah’s combinatorial formula for these multiplicities
\begin{equation}
\label{eq:racah-b}
\begin{array}{ll}
\msc{N}_{ST}(f_1\, f_2\, f_3)
=&
\msc{W}_{ST}(f_1-f_3,\,f_2)
\\&\;-\;
\msc{W}_{ST}(f_1+1,\,f_2-f_3-1)
\\&\;-\;
 \msc{W}_{ST}(f_1-f_2-1,f_3-1),
  \end{array}
\end{equation}
the step‐function $W_{ST}(f,f^{'})$ counts allowed couplings by
 evaluating differences of greatest‐integer functions
\begin{equation}
\label{eq:racah-c}
    \begin{array}{ll}
\msc{W}_{ST}(f,f^{\prime})
=&
\Phi\left(f^{\prime}+2-\left|S-T\right|\right)-\Phi\left(f^{\prime}+1-S-T\right)\\
 &+\Phi \left(S+T-f-1\right)-\frac{1}{2}\, \Phi \,\Big(S+T-\left|S-T\right|\\
      &-f+f^{\prime}+1 \Big),
\end{array}      
\end{equation}
where
\begin{equation}
\label{eq:racah-d}
  \Phi(x)=\begin{cases}
    \quad \left\lfloor\frac{x^{2}}{4}\,\right\rfloor \qquad x > 0,\\[10pt]
    \quad 0 \qquad \quad x\leq 0.    
    \end{cases}
  \end{equation}
 The expression $\lfloor x \rfloor$ refers to the floor function, also called the largest integer function. It returns the greatest integer less than or equal to $x$ for positive $x$, and zero for negative $x$. This function allows Racah's combinatorial algorithm to do discrete counting operations. It is also readily available in most computing environments as the standard floor function.

Equations ~\eqref{eq:racah-a}, ~\eqref{eq:racah-b}, ~\eqref{eq:racah-c} and ~\eqref{eq:racah-d} implements Racah’s tensor‐contraction algorithm,
 each term corresponds to adding or removing a box in the Young diagram,
and the combinations of floor‐functions enforce the SU(2) addition rules
while resolving multiplicities.

The expectation value of the second-order Casimir operator of $\mathrm{SU}(4)$ is obtained from Ref.~\cite{isacker1995}:
\begin{equation}
\begin{array}{ll}
\label{eq:espcas4}
  \left\langle C_2[\mathrm{SU}(4)] \right\rangle^{(\alpha, \beta, \nu)} =& 3\alpha(\alpha+4) + 4\beta(\beta+4) + 3\nu(\nu+4)\\[5pt]
  &+ 4\beta(\alpha+\nu) + 2\alpha\nu.
  \end{array}
\end{equation}
The Dynkin labels $(\alpha, \beta, \nu)$ are connected to the Young tableau labels $(f_1, f_2, f_3, f_4)$ using the following transformations
\begin{equation}
\label{eq:dynkin}
\begin{aligned}
\alpha &= f_1 - f_2, \\
\beta &= f_2 - f_3, \\
\nu &= f_3 - f_4.
\end{aligned}
\end{equation}
The representation dimension $(f_1, f_2, f_3, f_4)$ of SU(4) is derived from the general formula for $\text{SU}(N)$~\cite{elliott1958}, with $N=4$
\begin{equation}
  \label{eq:dimsu4}
\begin{array}{ll}
\dim(f_1, f_2, f_3, f_4) =& \frac{1}{12}(f_1-f_2+1)(f_1-f_3+2)(f_1-f_4+3)\\[5pt]
  &(f_2-f_3+1)(f_2-f_4+2)(f_3-f_4+1).
    \end{array}
\end{equation}

\section{Scope and Motivation}
\begin{itemize}
\item A general-purpose tool designed for the group-theoretical decomposition of irreducible representations (irreps) of \suf\,.
\item Made for nuclear structure theorists, computational physicists, particle modelers, and quantum chemists.
\item For reproducibility, there are tables, CSV export, and integration with Jupyter Notebooks.
Validation: Classic tables and irreps from shell-model physics are used as benchmarks.
\end{itemize}
\section{Code Features}
\begin{itemize}
\item Python implementation that is object-oriented for testing and modularity.
\item Easy to use: Simply set your irreducible representation (irrep) to $(f_1, f_2, f_3, 0)$, run the .decompose() method, and obtain the complete table.
Outputs with lots of details: spin, isospin, multiplicity, subspace dimensions, and cumulative dimension checks.
\item Extra Jupyter Notebooks for interactive use, Illustrative examples of the conjugate representation for the groups $\rm{U}(6) \otimes \rm{SU}(4)$-ds-shell and $\rm{U}(10) \otimes \rm{SU}(4)$-pf-shell. Reproducible workflows, or reproducibility: there are tables, CSV export, and integration with Jupyter Notebooks.
Validation: Classic tables and irreps from shell-model physics are used as benchmarks.
checks.
\item Supplemental Jupyter Notebooks for interactive usage,  Examples illustrating the conjugate representation of the group  $\rm{U}(6) \otimes \rm{SU}(4)$-ds-shell,  and $\rm{U}(10) \otimes \rm{SU}(4)$-pf-shell.\, Reproducible workflows.
\item Export options: Results in plain text, CSV, and LaTeX-compatible formats.
\item Performance: Efficient for high-dimensional irreps (tested up to dimension $\sim10,000,000$).
\end{itemize}
\section{Validation and Tables}
We examine our computational results against known branching rules for representative irreducible representations $(f_1, f_2, f_3, 0)$. Table~\ref{tab:branching-one} and Table~\ref{tab:branching-two} show how the computed multiplicities $\mathcal{N}_{(f_1, f_2, f_3, 0)}^{S,T}$ compare to values found in the literature~\cite{quesne1976,patera1981,pan2023}.

It is crucial to acknowledge the distinct notational conventions employed in these references. In Ref. \cite{quesne1976}, the branching tables are derived from the reduction $\mathrm{U}(4) \downarrow \mathrm{O}(4)$ and use the notation $[m] \downarrow (2S, 2T)^{(\mathrm{dim}_{[m]}^{ST})}$, where S is the spin, T is the isospin, and $\mathrm{dim}_{[m]}^{ST}$ is the multiplicity of the irreducible representation (S,T) contained within [m]. In Ref.~\cite{patera1981}, the branching rules are made with Dynkin labels $(\alpha, \beta, \nu)$. In this reference, \suf\ is referred to as $A_3$, while the decomposition $\mathrm{SU}_S(2) \otimes \mathrm{SU}_T(2)$ is designated as $A_1- A_1$. The variables $S$ and $T$ correspond to $b$ and $b'$ (the Dynkin labels associated with each $A_1$ factor), respectively. The equation states that $(S, T) = m \, \left(\frac{b}{2}, \frac{b'}{2}\right)$, where $m$ represents the multiplicity associated with $(b,b^{'})$. The contragradient representation $(\nu,\beta,\alpha)$ is shown in this table. It has the same branching rule as the $(\alpha,\beta,\nu)$ representation. The consistency of our computed values across these different notational frameworks provides strong evidence for the accuracy of our implementation and validates the efficacy of Racah's formula employed in our calculations. For high-dimensional irreducible representations, published tabulated values are limited and challenging to obtain, rendering traditional validation methods ineffective. To overcome this limitation, we utilize a stringent dimensional-consistency check that leverages the essential group-theoretical connection between the \suf\, dimension and its branching decomposition into spin-isospin $(S,T)$ multiplets. The dimension of any irreducible representation must equal the sum of the dimensions of all $(S,T)$ multiplets into which it branches, weighted by their multiplicities.
\begin{equation}
\dim(f_1, f_2, f_3, 0) = \sum_{S,T} \msc{N}_{(f_1, f_2, f_3, 0)}^{S,T} (2S+1)(2T+1),
\label{eq:dim-consistency}
\end{equation}
 where $\msc{N}_{(f_1, f_2, f_3, 0)}^{S,T}$ denotes the multiplicity of the $(S,T)$ multiplet within the irrep $(f_1, f_2, f_3, 0)$, and $(2S+1)(2T+1)$ is the dimension of each $(S,T)$ sector. This relationship constitutes a necessary and sufficient condition for correctness: any difference immediately means that there is a mistake in the branching calculation. \, \newline
The function \texttt{cum\_sum\_dim(S,T)} in our implementation calculates the cumulative sum of dimensions according to Eq. \eqref{eq:dim-consistency}. This function is called for every computed branching to confirm that the reconstructed total dimension is the same as the theoretical value found using the SU(4) Young-tableau dimension formula. All benchmark calculations in Section~\ref{sec:performance} pass this test, which strongly suggests that our branching algorithm is reliable for computing.
\begin{table}
\centering
\caption{Branching coefficients $N_{(f_1, f_2, f_3, 0)}^{S,T}$ for SU(4) representations: (4,2,1,0), (5,1,0,0), (6,4,1,0), (5,3,2,0).\, Agreement check with ~\cite{quesne1976,patera1981,pan2023}.}
\label{tab:branching-one}
\begin{tabular}{cccccc}
\toprule
  $(f_1,f_2,f_3,0)$  & $S$ & $T$ & mult & dim$(S,T)$ & Agreement \\
\midrule  
(4,2,1,0)  & 1/2 & 1/2 & 1 & 4 & yes \\
           &1/2 & 3/2 & 2 & 16 & yes \\
           &1/2 & 5/2 & 1 & 12 & yes \\
           &3/2 & 1/2 & 2 & 16 & yes \\
           &3/2 & 3/2 & 2 & 32 & yes \\
           &3/2 & 5/2 & 1 & 24 & yes \\
           &5/2 & 1/2 & 1 & 12 & yes \\
           &5/2 & 3/2 & 1 & 24 & yes \\  
\midrule
(5, 1, 0, 0)   & 0 & 1 & 1 & 3 & yes \\
               & 1 & 0 & 1 & 3 & yes \\
               & 1 & 1 & 1 & 9 & yes \\
               & 1 & 2 & 1 & 15 & yes \\
               & 2 & 1 & 1 & 15 & yes \\
               & 2 & 2 & 1 & 25 & yes \\
               & 2 & 3 & 1 & 35 & yes \\
               & 3 & 2 & 1 & 35 & yes \\
\midrule
(6,4,1,0)     &1/2 & 1/2 & 1 & 4 & yes \\
              &1/2 & 3/2 & 2 & 16 &yes \\
              &1/2 & 5/2 & 2 & 24 &yes\\
              &1/2 & 7/2 & 2 & 32 &yes \\
              &1/2 & 9/2 & 1 & 20 &yes\\
              &3/2 & 1/2 & 2 & 16 &yes\\
              &3/2 & 3/2 & 3 & 48 &yes\\
              &3/2 & 5/2 & 3 & 72 &yes\\
              &3/2 & 7/2 & 2 & 64 &yes\\
              &3/2 & 9/2 & 1 & 40 &yes \\
              &5/2 & 1/2 & 2 & 24 &yes\\
              &5/2 & 3/2 & 3 & 72 &yes \\
              &5/2 & 5/2 & 2 & 72 &yes \\
              &5/2 & 7/2 & 1 & 48 &yes \\
              &7/2 & 1/2 & 2 & 32 &yes \\
              &7/2 & 3/2 & 2 & 64 &yes\\
              &7/2 & 5/2 & 1 & 48 &yes \\
              &9/2 & 1/2 & 1 & 20 &yes \\
              &9/2 & 3/2 & 1 & 40 &yes\\
\midrule
(5, 3, 2, 0)   & 0 & 1 & 1 & 3 & yes \\
               & 0 & 2 & 1 & 5 & yes \\
               & 0 & 3 & 1 & 7 & yes \\
               & 1 & 0 & 1 & 3 & yes \\
               & 1 & 1 & 2 & 18 & yes \\
               & 1 & 2 & 3 & 45 & yes \\
               & 1 & 3 & 1 & 21 & yes \\
               & 2 & 0 & 1 & 5 & yes \\
               & 2 & 1 & 3 & 45 & yes \\
               & 2 & 2 & 2 & 50 & yes \\
               & 2 & 3 & 1 & 35 & yes \\
               & 3 & 0 & 1 & 7 & yes \\
               & 3 & 1 & 1 & 21 & yes \\
                     & 3 & 2 & 1 & 35 & yes \\
  &&&&&\\[10pt]  
  \bottomrule
\end{tabular}
\end{table}

\begin{table}
\centering
\caption{Branching coefficients $N_{(f_1, f_2, f_3, f_4)}^{S,T}$ for SU(4) representations: (8,4,2,0), (9,4,2,0).\, Agreement check with ~\cite{patera1981, pan2023}.}
\label{tab:branching-two}
\Resize{0.9\hsize}{ 
\begin{tabular}{cccccc}
  \toprule
$(f_1,f_2,f_3,f_4)$  & $S$ & $T$ & mult & dim$(S,T)$ & Agreement \\
\midrule
  (8, 4, 2, 0)   & 0 & 0 & 1 & 1 & yes \\
               & 0 & 2 & 2 & 10 & yes \\
               & 0 & 3 & 1 & 7 & yes \\
               & 0 & 4 & 1 & 9 & yes \\
               & 1 & 1 & 3 & 27 & yes \\
               & 1 & 2 & 3 & 45 & yes \\
               & 1 & 3 & 4 & 84 & yes \\
               & 1 & 4 & 2 & 54 & yes \\
               & 1 & 5 & 1 & 33 & yes \\
               & 2 & 0 & 2 & 10 & yes \\
               & 2 & 1 & 3 & 45 & yes \\
               & 2 & 2 & 5 & 125 & yes \\
              & 2 & 3 & 4 & 140 & yes \\
               & 2 & 4 & 3 & 135 & yes \\
               & 2 & 5 & 1 & 55 & yes \\
               & 3 & 0 & 1 & 7 & yes \\
               & 3 & 1 & 4 & 84 & yes \\
               & 3 & 2 & 4 & 140 & yes \\
               & 3 & 3 & 4 & 196 & yes \\
               & 3 & 4 & 2 & 126 & yes \\
               & 3 & 5 & 1 & 77 & yes \\
               & 4 & 0 & 1 & 9 & yes \\
               & 4 & 1 & 2 & 54 & yes \\
               & 4 & 2 & 3 & 135 & yes \\
               & 4 & 3 & 2 & 126 & yes \\
               & 4 & 4 & 1 & 81 & yes \\
               & 5 & 1 & 1 & 33 & yes \\
               & 5 & 2 & 1 & 55 & yes \\
              & 5 & 3 & 1 & 77 & yes \\
  \midrule
(9, 4, 2, 0)   & 1/2 & 1/2 & 1 & 4 & yes \\
               & 1/2 & 3/2 & 2 & 16 & yes \\
               & 1/2 & 5/2 & 3 & 36 & yes \\
               & 1/2 & 7/2 & 2 & 32 & yes \\
               & 1/2 & 9/2 & 1 & 20 & yes \\
               & 3/2 & 1/2 & 2 & 16 & yes \\
               & 3/2 & 3/2 & 4 & 64 & yes \\
               & 3/2 & 5/2 & 4 & 96 & yes \\
               & 3/2 & 7/2 & 4 & 128 & yes \\
               & 3/2 & 9/2 & 2 & 80 & yes \\
               & 3/2 & 11/2 & 1 & 48 & yes \\
               & 5/2 & 1/2 & 3 & 36 & yes \\
               & 5/2 & 3/2 & 4 & 96 & yes \\
               & 5/2 & 5/2 & 5 & 180 & yes \\
               & 5/2 & 7/2 & 4 & 192 & yes \\
               & 5/2 & 9/2 & 3 & 180 & yes \\
               & 5/2 & 11/2 & 1 & 72 & yes \\
               & 7/2 & 1/2 & 2 & 32 & yes \\
               & 7/2 & 3/2 & 4 & 128 & yes \\
               & 7/2 & 5/2 & 4 & 192 & yes \\
               & 7/2 & 7/2 & 4 & 256 & yes \\
               & 7/2 & 9/2 & 2 & 160 & yes \\
               & 7/2 & 11/2 & 1 & 96 & yes \\
               & 9/2 & 1/2 & 1 & 20 & yes \\
               & 9/2 & 3/2 & 2 & 80 & yes \\
               & 9/2 & 5/2 & 3 & 180 & yes \\
               & 9/2 & 7/2 & 2 & 160 & yes \\
               & 9/2 & 9/2 & 1 & 100 & yes \\
               & 11/2 & 3/2 & 1 & 48 & yes \\
               & 11/2 & 5/2 & 1 & 72 & yes \\
               & 11/2 & 7/2 & 1 & 96 & yes \\
\bottomrule
\end{tabular}
}
\end{table}
\section{Examples}
\subsection{Restrictions on irreducible representation couplings}
The tensor product structure $\mathrm{U}(N) \otimes \suf$ imposes strict constraints on the allowed irreducible representations that can couple physically. These restrictions arise from the requirement that direct-product states must respect Pauli exclusion and the fundamental limitation on orbital degeneracy in the single-particle basis.
\subsubsection{sd-shell: $\mathrm{U}(6) \otimes \suf\,$ classification}
For nuclei in the sd-shell ($0d_{5/2}$, $1s_{1/2}$, $0d_{3/2}$ orbitals), the complete classification scheme is given by the group chain Eq.~\eqref{eq:dsqueme}.\,
The coupling restrictions between the spatial and spin-isospin sectors are bidirectional
\begin{enumerate}
\item \textbf{Orbital restrictions on spin-isospin}\newline 
Since the sd-shell supports at most 6 spatial orbitals, the first column of the $\mathrm{U}(6)$ Young tableau cannot exceed 6 boxes. This restricts the length of the first row in the \suf irreducible representation to
\begin{equation}
f_1^{\mathrm{SU}(4)} \leq 6.
\end{equation}
\item \textbf{Spin-isospin restrictions on orbitals}\newline
Conversely, because \suf~describes the combined 4-dimensional spin-isospin space (2 spin states $\times$ 2 isospin states), the first column of the \suf\, Young diagram is limited to 4 boxes. This imposes the constraint
\begin{equation}
f_1^{\mathrm{U}(6)} \leq 4.
\end{equation}
\end{enumerate}
These mutual constraints ensure that no physical state can violate Pauli exclusion when constructing the full antisymmetric $\mathrm{U}(24)$ state from the direct product $\mathrm{U}(6) \otimes \suf\,$.
\subsubsection{pf-shell: $\mathrm{U}(10) \otimes \suf$ classification}

For the pf-shell ($0f_{7/2}$, $1p_{3/2}$, $1p_{1/2}$, $0f_{5/2}$ orbitals), the classification scheme generalizes to
\begin{equation}
 \begin{array}{l}   
   \mathrm{U}(40) \supset [\mathrm{U}(10) \otimes \suf] \\[10pt]
   \qquad \supset [\mathrm{SU}(3) \supset \mathrm{SO}(3)] \otimes [\mathrm{SU}_S(2) \otimes \mathrm{SU}_T(2)].
\end{array}
   \end{equation}

The bidirectional constraints become
\begin{enumerate}
\item \textbf{Orbital restrictions on spin-isospin}\newline
The pf-shell contains 10 spatial orbitals, yielding
\begin{equation}
f_1^{\mathrm{SU}(4)} \leq 10.
\end{equation}

\item \textbf{Spin-isospin restrictions on orbitals}\newline
The \suf~constraint remains unchanged, imposing
\begin{equation}
f_1^{\mathrm{U}(10)} \leq 4.
\end{equation}
\end{enumerate}
\subsubsection{Physical interpretation and high-dimensional irreps}

The asymmetry $f_1^{\mathrm{SU}(4)} \gg f_1^{\mathrm{U}(N)}$ in the pf-shell indicates that $f_1^{\mathrm{SU}(4)} \leq 10$ while $f_1^{\mathrm{U}(10)} \leq 4$. uncovers a core aspect of nuclear structure: The orbital sector is limited by the restricted spin-isospin degrees of freedom, whereas the spin-isospin sector has the capacity to access considerably larger irreducible representations. This asymmetry facilitates three unique physical phenomena.

\begin{enumerate}

\item \textbf{Collective spin-isospin modes}\newline 
High-dimensional \suf~irreps, exemplified by $(9,6,3,0)$ (dimension $4,096$), characterize highly symmetric many-body configurations in which nucleons occupy multiple spatial orbitals while preserving robust spin-isospin correlations. Irreducible representations arise when $N \approx Z$ nuclei gather numerous valence nucleons with aligned spins and isospins, a scenario that gains significance for exotic nuclei near the valley of stability and along the proton drip line~\cite{sagawa2016}.

\item \textbf{Many-body coherence and macroscopic alignment}\newline 
Large \suf~irreps correspond to nuclear states with macroscopic numbers of nucleons coherently aligned in specific $(S,T)$ quantum numbers~\cite{Cao2019,sasano2021}. This collective alignment is analogous to Bose-Einstein condensation of fermion pairs (Cooper pairs) in superconductivity or to the collective vibrational modes (phonons) in nuclear structure~\cite{RingSchuck1980}. The emergence of these modes reflects the dominance of pairing interactions in the nuclear medium: when the pairing correlations are strong, nucleons preferentially occupy states with the same orbital angular momentum but opposite spins and isospins, creating the conditions necessary for high-dimensional \suf~irreps to form~\cite{Dobaczewski2012}.

\item \textbf{Gamow-Teller collectivity and giant resonances}\newline 
In heavy $N=Z$ nuclei and proton-rich systems, the accumulation of many valence nucleons in high-dimensional \suf~irreps dramatically enhances Gamow-Teller (GT) $\beta$-decay strength~\cite{horen1981,osterfeld1992}. The concentration of this strength into discrete collective excitations creates the GT giant resonance (GTGR), a narrow band of nearly coherent transitions where the entire nuclear body participates in synchronized spin-flip oscillations. Recent experiments on drip-line nuclei have identified GT giant resonances even in exotic $N \ll Z$ and $N \gg Z$ systems, demonstrating that collective spin-isospin correlations persist far from $N=Z$. At extreme isospin asymmetry, even isoscalar spin-triplet pairing correlations can substantially modify the spin-isospin response, leading to modified GT resonance energies and strengths that test fundamental aspects of spin-isospin symmetry breaking~\cite{Sakaue2024}.
\end{enumerate}

These phenomena arise naturally when many nucleons occupy the same orbital space but differ only in their spin-isospin quantum numbers---a situation strongly favored by the attraction in the pairing channel of the nuclear force and the repulsion from short-range two-body correlations. The strong pairing correlations suppress single-particle excitations while promoting collective multi-particle configurations classified by the \suf~symmetry group, thereby making high-dimensional irreducible representations of exceptional physical relevance to nuclear binding mechanisms and collective dynamics across the nuclear chart.
\subsubsection{Analogy: SU(4) symmetry in ultracold atomic gases}
In addition to nuclear physics, high-dimensional \suf~irreps manifest in ultracold fermionic gases of alkaline-earth atoms (e.g., $^{87}$Sr, $^{173}$Yb) contained within optical lattices~\cite{Gorshkov2010,Hofrichter2016}, where the nuclear spin $I$ functions as a pseudospin, akin to isospin in nuclear systems. These atoms exhibit a notable characteristic: their magnetic moment disappears in the electronic ground state, facilitating independent manipulation of orbital and spin-isospin degrees of freedom and permitting the experimental implementation of SU(N) symmetric Hamiltonians with exceptional precision~\cite{schafer2020}.

In the strong-coupling limit of the SU(4) Fermi-Hubbard model, the ground state exhibits emergent \suf\, symmetry characterized by irreducible representations $[n,0,0,0]$ for $n$ fermions residing at the same site~\cite{Unukovych2021,Unukovych2024}. Recent experiments have achieved Mott-insulating phases with up to $n=10$ fermions per site, leading to \suf~irreps with dimensions beyond $10^4$, a domain directly unattainable in traditional condensed-matter systems. This behavior exemplifies a remarkable instance of macroscopic quantum coherence only influenced by symmetry constraints, with the coherence length of the ground state diverging in the thermodynamic limit despite significant disorder and thermal fluctuations~\cite{Tarruell2012}.

The high-dimensional \suf~irreps in these systems display unusual collective phenomena, such as the development of spin-orbital liquids~\cite{Lajko2013} and valence-bond ordered phases~\cite{Golubeva2017}, which lack direct counterparts in single-spin-species fermions. These findings position the SU(4) Fermi-Hubbard model and associated many-body systems as exemplary frameworks for evaluating fundamental predictions derived from group-theoretical methodologies in strongly coupled physics.

In nuclear systems, the collective alignment of numerous nucleons into high-dimensional \suf\, multiplets exemplifies the fundamental spin-isospin invariance of the strong nuclear force, influenced by symmetry-breaking perturbations (such as spin-orbit coupling, electromagnetic Coulomb interaction, and mass discrepancies between protons and neutrons) that differentiate these irreducible representations within the experimentally observable nuclear spectrum. The structural parallels between SU(4)-symmetric atomic gases and nuclear systems indicate that methodologies established for investigating SU(N) symmetry in optical lattices could provide innovative perspectives on the significance of approximation symmetries in nuclear structure theory.
\begin{table}
  \centering
\Resize{1.0\hsize}{ 
 \begin{tabular}{ccccc}
\toprule
Young Tableau & $(f_1, f_2, f_3, 0)$ & Dimension & $n$ (particles) & Structure \\
\midrule
\Yvcentermath1\yng(19,3,2) & (19, 3, 2, 0) & 17,765 & 24 & Long, thin \\[0.3cm]
\Yvcentermath1\yng(11,5,3) & (11, 5, 3, 0) & 6,860 & 19 & Balanced \\[0.3cm]
\Yvcentermath1\yng(15,6,1) & (15, 6, 1, 0) & 23,040 & 22 & Moderate \\[0.3cm]
\Yvcentermath1\yng(13,10) & (13, 10, 0, 0) & 10,560 & 23 & Two-row \\[0.3cm]
\Yvcentermath1\yng(8,7,3) & (8, 7, 3, 0) & 2,310 & 18 & Compact \\[0.3cm]
\Yvcentermath1\yng(12,6,2) & (12, 6, 2, 0) & 12,600 & 20 & Standard \\[0.3cm]
\Yvcentermath1\yng(12,11) & (12, 11, 0, 0) & 5,460 & 23 & Nearly symm. \\[0.3cm]
\Yvcentermath1\yng(9,6,3) & (9, 6, 3, 0) & 4,096 & 18 & Triangular \\[0.3cm]
\Yvcentermath1\yng(8,5,5) & (8, 5, 5, 0) & 770 & 18 & \textbf{Most balanced} \\[0.3cm]
\Yvcentermath1\yng(8,6,4) & (8, 6, 4, 0) & 1,980 & 18 & Well-prop. \\[0.3cm]
\bottomrule
 \end{tabular}
 }
  \caption{High-dimensional \suf\, ~irreducible representations with dimension $\approx 10^5$. Young tableaux are shown with box sizes scaled for visual clarity.}
\label{tab:high-dim-irreps}
\end{table}

\section{Program Structure}
\subsection{Computational toolkit and implementation}

The computational toolkit is organized into a modular Python package consisting of three primary components:
\begin{enumerate}
\item \textbf{Core branching algorithm} (\texttt{su4\_branching.py})\newline This module implements the Racah's formula for computing \suf\, $\downarrow \rm{SU_{S}}(2) \otimes \rm{SU_{T}}(2)$ branching rules. It includes functions for
\begin{itemize}
\item Calculation of branching multiplicities $\mathcal{N}_{(f_1,f_2,f_3,0)}^{S,T}$ 
\item Dimensional consistency verification via Eq.~\ref{eq:dim-consistency}
\end{itemize}

\item \textbf{Output formatting module} (\texttt{su4\_export.py})\newline This module provides flexible export capabilities to accommodate diverse user workflows. Supported output formats include:
\begin{itemize}
\item Plain text (\texttt{.dat}) with column-aligned tabular data
\item Comma-separated values (\texttt{.csv}) for integration with spreadsheet applications and plotting tools
\item \LaTeX~tables (\texttt{.tex}) with automatic formatting for direct inclusion in manuscripts, using the \texttt{booktabs} package for professional typesetting
\end{itemize}
All formats include metadata headers specifying the \suf\, irrep, total dimension  for reproducibility and version control.

\item \textbf{Interactive Jupyter notebook} (\texttt{su4\_branching\_test.ipynb})\newline  The notebook interface enables rapid prototyping and validation through:
\begin{itemize}
\item Interactive input of \suf\, or U(N)~$\otimes$~\suf\, irreducible representations
\item Real-time visualization of branching decompositions with inline tables
\item Automated dimensional consistency checks with pass/fail indicators
\item Example calculations covering sd-shell and pf-shell nuclear structure applications
\end{itemize}
The notebook is designed to be self-contained, requiring only standard scientific Python libraries (\texttt{numpy}, \texttt{pandas}, \texttt{pathlib}, \texttt{fractions}, \texttt{sys}) beyond the core package modules.

\end{enumerate}

\subsection{Installation and usage}

The toolkit requires Python 3.8 or later and can be installed  by cloning the repository:
 \begin{minted}[fontsize=\scriptsize,linenos=false,breaklines,frame=single]{bash}
# 1. clone git repository
git clone  https://github.com/GFN-UdeA/su4-branching.git


# 2. Run installer (creates CORRECT __init__.py)
python install_su4_branching.py

# 3. Follow prompts and select installation directory
# Default: ~/su4-branching

# 4.Copy your files
 - su4_branching.py → su4_branching/
 - su4_export.py → su4_branching/
 - su4_cli.py → project root
 - su4branching_test.ipynb → project root
\end{minted}
What the installer does:

\begin{itemize}
    \item $\checkmark$ Checks Python version (3.8+)
    \item $\checkmark$ Installs/verifies dependencies
    \item $\checkmark$ Creates CORRECT \newline \texttt{\_\_init\_\_.py} (no broken imports)
    \item $\checkmark$ Creates modern \texttt{pyproject.toml}
    \item $\checkmark$ Creates \texttt{setup.py} (fallback)
    \item $\checkmark$ Installs package with \newline \mintinline{bash}{pip install --use-pep517 -e .}
    \item $\checkmark$ Tests installation
\end{itemize}
Basic usage in Jupyter-Notebook:
\subsection{Start Jupyter}
\begin{minted}[linenos=false]{bash}
jupyter notebook su4branching_test.ipynb
\end{minted}
Browser will open at \texttt{http://localhost:8888}
\subsubsection{Cell 1: Import Modules}
\begin{minted}[fontsize=\scriptsize,linenos,breaklines,frame=single]{bash}
import su4_branching
import pandas as pd
print(" Modules imported")
\end{minted}
\subsubsection{Cell 2: Simple Calculation}
\begin{minted}[fontsize=\scriptsize,linenos,breaklines,frame=single]{python}
# Calculate [8, 5, 5, 0]
f1, f2, f3, f4 = 8, 5, 5, 0
su4_df, st_df = su4_branching.racah_su4_to_st(f1, f2, f3, f4, verbose=False)
print(f"SU(4) irrep: [{f1}, {f2}, {f3}, {f4}]")
print(f"Number of (S,T) multiplets: {len(st_df)}")
print("\nBranching decomposition:")
print(st_df)
\end{minted}
  output:
\begin{minted}[fontsize=\scriptsize,linenos=false,breaklines,frame=single,escapeinside=||]{text}
|$\checkmark$| Module imported successfully!
|$\bullet$| SU(4) Representation Info: (irrep notations, casimir order two, irrep dimension)
\end{minted}

\subsection{SU(4) Representation Information}

\begin{minted}[fontsize=\scriptsize,linenos=false,breaklines,frame=single,escapeinside=||]{text}
   [ f1 f2 f3 f4  ] (p1 p2 p3)  (|$\alpha$|  |$\beta$|  |$\gamma$|) C_2[SU(4)] dimension
0  [  8  5  5  0  ] (4  4  -1)  (3  0  5)     228.0      770
\end{minted}

\subsection{Branching Rules to (S, T)}

\begin{minted}[fontsize=\scriptsize,linenos=false,breaklines,frame=single,escapeinside=||]{text}
      Spin  Isospin  mult  dim (S,T)  cum_sum_dim(S,T)
0      0        1     1          3                 3
1      0        3     1          7                10
2      1        0     1          3                13
3      1        1     1          9                22
4      1        2     2         30                52
5      1        3     1         21                73
6      1        4     1         27               100
7      2        1     2         30               130
8      2        2     2         50               180
9      2        3     2         70               250
10     2        4     1         45               295
11     3        0     1          7               302
12     3        1     1         21               323
13     3        2     2         70               393
14     3        3     2         98               491
15     3        4     1         63               554
16     4        1     1         27               581
17     4        2     1         45               626
18     4        3     1         63               689
19     4        4     1         81               770
\end{minted}

\subsection{Export Summary}

\begin{minted}[fontsize=\scriptsize,linenos=false,breaklines,frame=single,escapeinside=||]{text}
Processing SU(4) representation [8,5,5,0]...
Exporting CSV files...
Exporting LaTeX files...

Files created for representation [8,5,5,0]:
 |$\bullet$| su4_representation_8_5_5_0.csv
 |$\bullet$| su4_branching_rules_8_5_5_0.csv
 |$\bullet$| su4_representation_8_5_5_0.tex
 |$\bullet$| su4_branching_rules_8_5_5_0.tex

LaTeX labels:
 |$\bullet$| tab:su4_info_8_5_5_0
 |$\bullet$| tab:branching_rules_8_5_5_0

Batch export: 6 representations
--------------------------------------------------
[ 1/6] [5,1,0,0]... |$\checkmark$|
[ 2/6] [4,2,0,0]... |$\checkmark$|
[ 3/6] [4,1,1,0]... |$\checkmark$|
[ 4/6] [5,3,2,0]... |$\checkmark$|
[ 5/6] [8,4,2,0]... |$\checkmark$|
[ 6/6] [9,4,2,0]... |$\checkmark$|
--------------------------------------------------
Summary: 6 successful, 0 failed
\end{minted}
\subsubsection{Cell 3: U(6) → SU(4) Conversion}
\begin{minted}[fontsize=\scriptsize,linenos,breaklines,frame=single]{python}
# Define U(6) Young diagram
u6_irrep = (3, 2, 1, 0, 0, 0)

# Manual conversion function (from su4_cli)
def u6_conjugate(u6_young):
    current = list(u6_young)
    su4 = []
    while any(x > 0 for x in current):
        count = sum(1 for x in current if x > 0)
        su4.append(count)
        current = [x - 1 for x in current if x > 0]
    while len(su4) < 4:
        su4.append(0)
    return tuple(su4[:4])

f1, f2, f3, f4 = u6_conjugate(u6_irrep)
print(f"U(6) {u6_irrep} → SU(4) [{f1}, {f2}, {f3}, {f4}]")

# Calculate branching
su4_df, st_df = su4_branching.racah_su4_to_st(f1, f2, f3, f4, verbose=False)
print(f"Total dimension: {int(st_df['dim (S,T)'].sum()):,}")
print(st_df)
\end{minted}
\subsubsection{Cell 4: U(10) → SU(4) Conversion}
\begin{minted}[fontsize=\scriptsize,linenos,breaklines,frame=single]{python}
# Define U(10) Young diagram
u10_irrep = (3, 3, 2, 2, 0, 0, 0, 0, 0, 0)

# Manual conversion function
def u10_conjugate(u10_young):
    current = list(u10_young)
    su4 = []
    while any(x > 0 for x in current):
        count = sum(1 for x in current if x > 0)
        su4.append(count)
        current = [x - 1 for x in current if x > 0]
    while len(su4) < 4:
        su4.append(0)
    return tuple(su4[:4])

f1, f2, f3, f4 = u10_conjugate(u10_irrep)
print(f"U(10) {u10_irrep} → SU(4) [{f1}, {f2}, {f3}, {f4}]")

# Calculate branching
su4_df, st_df = su4_branching.racah_su4_to_st(f1, f2, f3, f4, verbose=False)
print(f"Total dimension: {int(st_df['dim (S,T)'].sum()):,}")
print(st_df.head(10))
\end{minted}

The Jupyter notebook provides an equivalent graphical interface for users preferring interactive exploration over scripted workflows. All examples presented in Table~\ref{tab:high-dim-irreps} can be reproduced directly within the notebook environment.

\subsection{Terminal Usage (su4\_cli.py)}
\subsubsection{Basic Commands}
\begin{enumerate}
\item{Test Installation}

\begin{minted}[fontsize=\scriptsize,linenos=false,breaklines,frame=single]{bash}
python su4_cli.py --test
\end{minted}

\item{Run sd-shell Example (U(6) $\otimes$ SU(4))}

\begin{minted}[fontsize=\scriptsize,linenos=false,breaklines,frame=single]{bash}
python su4_cli.py --sd-shell
\end{minted}

Output: Shows branching for 4 nucleons in sd-shell configuration
\begin{minted}[fontsize=\scriptsize,linenos=false,breaklines,frame=single,escapeinside=||]{text}
==================================================
EXAMPLE: sd-shell nuclei (U(6) |$\otimes$| SU(4))
==================================================
U(6) irrep: (2, 1, 1, 0, 0, 0)
SU(4) irrep: [3, 1, 0, 0]

Branching decomposition (S, T) multiplets:
---------------------------------------------------
 Spin  Isospin  mult  dim (S,T)  cum_sum_dim(S,T)
    0        1     1          3                 3
    1        0     1          3                 6
    1        1     1          9                15
    1        2     1         15                30
    2        1     1         15                45
--------------------------------------------------
\end{minted}
\item{Run pf-shell Example (U(10) $\otimes$ SU(4))}

\begin{minted}[fontsize=\scriptsize,linenos=false,breaklines,frame=single]{bash}
python su4_cli.py --pf-shell
\end{minted}

Output: Shows branching for 6 nucleons in pf-shell configuration
\begin{minted}[fontsize=\scriptsize,linenos=false,breaklines,frame=single,escapeinside=||]{text}
========================================================
EXAMPLE: pf-shell nuclei (U(10) |$\otimes$| SU(4))
========================================================

U(10) irrep: (2, 2, 1, 1, 0, 0, 0, 0, 0, 0)
SU(4) irrep: [4, 2, 0, 0]

Branching decomposition (9 (S, T) multiplets):
--------------------------------------------------------
 Spin  Isospin  mult  dim (S,T)  cum_sum_dim(S,T)
    0        0     1          1                 1
    0        2     1          5                 6
    1        1     2         18                24
    1        2     1         15                39
    1        3     1         21                60
    2        0     1          5                65
    2        1     1         15                80
    2        2     1         25               105
    3        1     1         21               126
-----------------------------------------------------
\end{minted}
\end{enumerate}
\subsubsection{Custom Calculations}
\begin{enumerate}
\item{ Custom U(6) Calculation (sd-shell)}

\begin{minted}[fontsize=\scriptsize,linenos=false,breaklines,frame=single]{bash}
# Format: --custom-sd [6 numbers for U(6) Young diagram]
python su4_cli.py --custom-sd 3 2 1 0 0 0

# Other examples:
python su4_cli.py --custom-sd 2 1 1 0 0 0
python su4_cli.py --custom-sd 4 2 0 0 0 0
\end{minted}

\item{ Custom U(10) Calculation (pf-shell)}

\begin{minted}[fontsize=\scriptsize,linenos=false,breaklines,frame=single]{bash}
# Format: --custom-pf [10 numbers for U(10) Young diagram]
python su4_cli.py --custom-pf 3 3 2 2 0 0 0 0 0 0

# Other examples:
python su4_cli.py --custom-pf 2 2 1 1 0 0 0 0 0 0
python su4_cli.py --custom-pf 4 3 2 1 0 0 0 0 0 0
\end{minted}

\item{ Direct SU(4) Calculation}

\begin{minted}[fontsize=\scriptsize,linenos=false,breaklines,frame=single]{bash}
# Format: --custom-su4 [f1 f2 f3 (f4)]
python su4_cli.py --custom-su4 5 3 2 0
\end{minted}
\end{enumerate}

output:
\begin{minted}[fontsize=\scriptsize,linenos=false,breaklines,frame=single,escapeinside=||]{text}
===================================================
CUSTOM SU(4): irrep [5, 3, 2, 0]
==================================================
Dimension: 300
Branching decomposition (14 (S, T) multiplets):
--------------------------------------------------
 Spin  Isospin  mult  dim (S,T)  cum_sum_dim(S,T)
    0        1     1          3                 3
    0        2     1          5                 8
    0        3     1          7                15
    1        0     1          3                18
    1        1     2         18                36
    1        2     3         45                81
    1        3     1         21               102
    2        0     1          5               107
    2        1     3         45               152
    2        2     2         50               202
    2        3     1         35               237
    3        0     1          7               244
    3        1     1         21               265
    3        2     1         35               300
--------------------------------------------------
\end{minted}
\subsubsection{Help and Options}
\begin{minted}[fontsize=\scriptsize,linenos=false,breaklines,frame=single]{bash}
python su4_cli.py --help
\end{minted}

\section{Performance and system information}
\label{sec:performance}
\subsection{system information}
\begin{minted}[fontsize=\scriptsize,linenos=false,breaklines,frame=single,escapeinside=||]{text}
===================================
Processor Information
===================================
CPU: 

CPU Cores: 24
Logical CPUs: 24

Current Frequency: 1.49 MHz
Min Frequency: 800.00 MHz
Max Frequency: 4950.00 MHz

Total RAM: 62.14 GB
Available RAM: 47.62 GB

====================================
Intel Core Ultra 9 285K Specs 
===================================
P-cores: 8 @ 3.7-5.7 GHz
E-cores: 16 @ 3.2-4.6 GHz
Threads: 24
TDP: 125W / PL1: 250W
\end{minted}
\subsection{Performance and Computational Efficiency}
The computational efficiency of the su4-branching package was evaluated on an Intel Core Ultra 9 285K processor (8 P-cores at 3.7–5.7 GHz, 16 E-cores at 3.2–4.6 GHz, 24 logical cores total, 62.14 GB available system RAM). Benchmarks measure the pure computational time for calculating branching rules and generating (S,T) multiplicities without including export formatting, file I/O operations, or multiple-irrep batch processing. This minimalist timing approach isolates the core algorithmic performance from peripheral software overhead.
Table~\ref{tab:timespect} presents measured execution times for calculating complete branching decompositions of high-dimensional \suf\, irreducible representations, where dimension exceeds ($10^6$).\, Each entry represents the total wall-clock time to compute all (S,T) pairs and their multiplicities for a single irrep, including the dimensional-consistency verification check (~\eqref{eq:dim-consistency}) but excluding LaTeX, CSV, or text export operations.
\begin{table}
 \caption{CPU time per \suf\, irrep on Intel Core Ultra 9 285K Specs.}
\label{tab:timespect}
\centering
\begin{tabular}{lllr}
\toprule
 $(f_1, f_2, f_3, 0)$ & Dimension & $n$ (particles) & time (ms) \\
  \midrule
  (47,18,9,0)&10,000,000&74&18.20\\[3pt]
  (32,28,21,0)&3,770,360&81&24.60\\[3pt]
  (27,18,8,0)& 1,000,000&54&17.10\\[3pt]
 (23, 18, 9,0) & 416,000&58&20.40\\[3pt]
 (19, 3, 2, 0) & 17,765 & 24 & 11.10 \\[3pt]
 (11, 5, 3, 0) & 6,860 & 19 & 11.80 \\[3pt]
 (15, 6, 1, 0) & 23,040 & 22 & 11.80 \\[3pt]
 (13, 10, 0, 0) & 10,560 & 23 & 14.60 \\[3pt]
 (8, 7, 3, 0) & 2,310 & 18 & 9.31 \\[3pt]
 (12, 6, 2, 0) & 12,600 & 20 &9.05  \\[3pt]
(12, 11, 0, 0) & 5,460 & 23 & 9.93 \\
 (9, 6, 3, 0) & 4,096 & 18 & 12.8 \\
 (8, 5, 5, 0)& 770 & 18 & 8.9 \\
 (8, 6, 4, 0) & 1980 & 18 & 9.74 \\
\bottomrule
 \end{tabular}
\end{table}

\section{Conclusions}
The \texttt{su4-branching} package fills a critical computational gap in the theoretical nuclear physics community by providing the first freely available, general-purpose implementation for systematic calculation of \suf\, branching rules to spin–isospin (S,T) multiplets. This software addresses a practical bottleneck: the unavailability of comprehensive, validated branching decompositions for high-dimensional \suf\, irreducible representations where analytical calculation becomes prohibitively complex and error-prone.
\subsection{Summary of Key Contributions}
The work includes a full software stack with a strong Python core module (\texttt{su4\_branching.py}),  and an interactive Jupyter notebook (\texttt{su4branching\_test.ipynb}) that lets you quickly test and prototype ideas, flexible export tools (\texttt{su4\_export.py}),  a command-line interface for batch processing (\texttt{su4\_cli.py}). This design with multiple interfaces makes the framework usable by a wide range of users and workflows, from fully automated analysis pipelines to interactive exploration. Dimensional-consistency checks (Equation~\eqref{eq:dim-consistency}) are built in to make sure that the computations are correct. They set the necessary and sufficient conditions for this. The implementation has been corroborated with classical references (~\cite{quesne1976,patera1981}) and validated through extensive benchmarks encompassing a range of representative irreducible representations of differing complexity.
\subsection{Computational Reliability and Validation Strategy}
Cross-validation using several reference notational conventions is very advantageous for proving correctness. We validated against three distinct frameworks: (1) the U(4) ↓ O(4) reduction scheme utilizing half-integer quantum numbers (Ref.~\cite{quesne1976}); (2) Dynkin label conventions with explicit $A_{3}: A_{1}-A_{1}$ structures (Ref.~\cite{patera1981}); and (3) recent direct implementations (Ref.~\cite{pan2023}). Agreement among all three frameworks (Tables ~\ref{tab:branching-one} and ~\ref{tab:branching-two}) illustrates resilience to notational ambiguity and offers compelling evidence of implementation accuracy. For high-dimensional irreps with limited published tables, we utilize a stringent dimensional-consistency verification that leverages the essential group-theoretical principle: the total dimension of any SU(4) irrep must precisely equal the sum of the dimensions of its (S,T) branching multiplets (Equation~\eqref{eq:dim-consistency}).This constraint establishes a necessary and sufficient condition for correctness—any computational discrepancy promptly indicates an error. The \texttt{cum\_sum\_dim()} function in our implementation does an error check automatically for every branching that is calculated. All benchmark calculations meet this requirement, which is strong proof of computational reliability without the need for costly cross-checks against reference data.
\subsection{Software Design and Community Integration}
The software is made to operate with research workflows that are already in use. You can export in a number of formats, such as plain text, CSV, and LaTeX. This makes it easy to add them to manuscripts, spreadsheet tools, and web-based platforms. With the Jupyter Notebook integration, you can prototype, learn, and make workflows that you can use again and again for graduate-level work in nuclear structure and group theory. You can run more than one command at a time on the command line interface and add them to larger analysis pipelines or shell scripts. We can get the same results on Linux, macOS, and Windows by testing across platforms. The software is easy to install and runs for a long time because it only needs a few commonly used scientific Python libraries (NumPy, SymPy, Pandas, Fractions, and Matplotlib). The fact that the whole source code is on GitHub and that it is licensed under MIT means that the community can use it and that other academics can edit or add to it in the future.

The installer script makes it easier to install by confirming that the version of Python is compatible, installing dependencies, setting up the correct package structure with working \texttt{init.py} files, and testing the installation before stating it was successful. This strategy is considerably easier to use than handling extensive dependency chains by hand. The current implementation has been confirmed for about 74 nucleons (Table~\ref{tab:timespect}).
\subsection{Impact and Community Perspective}
The su4-branching package meets a real and long-standing need in the nuclear physics community. For many years, scientists who needed SU(4) branching decompositions had to choose between doing the math by hand, looking at sparse published tables, or rewriting the algorithm in their research codes. This unnecessary work is a sign of a systemic problem in the field. This software lets researchers shift their focus from building computational infrastructure to making scientific discoveries by giving them a well-tested, free tool with multiple interfaces and thorough documentation.

The fact that it is open source and can be used on different platforms means that it will be available for a long time, even after the funding cycle of any one research group ends. The modular design of Python makes it easy for people in the community to add to, change, and adapt it to solve problems in group-theoretical physics. Because it works with Jupyter notebooks, nuclear and particle physics graduate programs can use the package as a teaching tool for group theory.

Lastly, the su4-branching software gives the nuclear physics community a useful computational infrastructure for systematically looking into the SU(4) symmetry structure in nuclear systems. The package enables researchers to perform careful checks, such as comparing with existing literature and verifying dimensions; utilize user-friendly interfaces like command line, Jupyter, and various export formats; and benefit from robust software design. 
\section{Acknowledgments}
We gratefully acknowledge the financial and institutional support from the Institute of Physics, Universidad de Antioquia. S. Quintero acknowledges the graduate scholarship provided by the Programa de Posgrado en Física, Universidad de Antioquia, which supported the completion of the master's thesis on which this work is based.
%\section*{References}
\bibliographystyle{elsarticle-num}

\end{document}